\documentclass[prd,aps,preprintnumbers,showpacs,10pt]{revtex4}
\usepackage{graphicx}
\usepackage{epstopdf}
\usepackage{color}
\usepackage{epsfig}
\renewcommand{\arraystretch}{1.8}

\newcommand{\be}{\begin{equation}}
\newcommand{\ee}{\end{equation}}
\newcommand{\bea}{\begin{eqnarray}}
\newcommand{\eea}{\end{eqnarray}}

\newcommand{\dd}{\displaystyle}

\def\slash#1{\setbox0=\hbox{$#1$}#1\hskip-\wd0\dimen0=5pt\advance
\dimen0 by-\ht0\advance\dimen0 by\dp0\lower0.5\dimen0\hbox
to\wd0{\hss\sl/\/\hss}} 
\begin{document}
\begin{titlepage}
\preprint{BARI-TH 584-07}
\title{AdS-QCD quark-antiquark potential, meson spectrum and tetraquarks}
\author{\textbf{M. V. Carlucci}}\author{\textbf{F. Giannuzzi}}
\author{\textbf{G. Nardulli}}
\author{\textbf{M. Pellicoro}}
\author{\textbf{S. Stramaglia}}
\affiliation{Universit\`a di Bari, I-70126 Bari, Italia}
\affiliation{I.N.F.N., Sezione di Bari, I-70126 Bari, Italia}
\begin{abstract}
AdS/QCD correspondence predicts the structure of quark-antiquark
potential in the static limit. We use this piece of information
together with the Salpeter equation (Schr\"odinger equation with
relativistic kinematics) and a short range hyperfine splitting
potential to determine quark masses and the quark potential
parameters from the meson spectrum. The agreement between theory and
experimental data is satisfactory provided one considers only mesons
comprising at least one heavy quark. We use the same potential (in
the one-gluon-exchange approximation) and these data to estimate
constituent diquark masses. Using these results as an input we
compute tetraquark masses using a diquark-antidiquark model. The
masses of the states $X(3872)$ or $Y(3940)$ are predicted rather
accurately. We also compute tetraquark masses with open charm and
 strangeness. Our result is that tetraquark candidates such
as $D_s(2317)$, $D_s(2457)$ or $X(2632)$ can hardly be interpreted
as diquark-antidiquark states within the present approach.
\end{abstract}

 \maketitle
\end{titlepage}
\setcounter{page}{1}%
\section{Introduction}Hadron spectroscopy has received in the last
few years a renewed attention, both experimental and theoretical. On
the experimental side several new charmonium and charmed states have
been observed (for reviews see e.g.
\cite{Ricciardi:2007rs,Zupanc:2007rw}).  The interpretation of the
new states has triggered  a considerable amount of theoretical work,
especially so because some of the new states cannot be interpreted
as ordinary quark-antiquark mesons (for reviews see
\cite{Swanson:2006st,Jaffe:2004ph}). For some of the new states,
such as the $X(3872)$ or $Y(3940)$
 \cite{Choi:2003ue,Abazov:2004kp,Acosta:2003zx,Aubert:2004ns,Abe:2005ix,Abe:2005iya,Abe:2004zs}
  an interpretation in terms of exotica
  (diquark-antidiquark bound states) has been given
  \cite{Maiani:2004uc,Maiani:2004vq,Maiani:2005pe,Maiani:2007vr},
   using a refined version of the constituent quark model
  \cite{De Rujula:1975ge}. On the other hand also more conventional
interpretations have been proposed. It is therefore useful to have
an independent approach to the calculation of tetraquark masses in
order to assess the validity of the diquark-antidiquark
interpretation of the new states. This is one of the aims of the
present paper. We approach this problem using a semirelativistic
method based on the use of a relativistic wave equation, the
Salpeter equation. This is a Schr\"odinger equation with the
relativistic kinematics (for previous use of this equation for
hadron spectroscopy see e.g. \cite{Colangelo:1990rv} and references
therein); as such, it  has the usual limitations of the static
potential approach, but presents the advantage of the relativistic
kinematics.

A crucial point in this approach is the choice of the potential.
There are several proposals in the literature, e.g. the Cornell
potential \cite{Eichten:1978tg} or its variants
\cite{Richardson:1978bt,Buchmuller:1980su}. We shall use here a
modified version of a static quark potential that has been recently
found in the context of the AdS/QCD correspondence
\cite{Andreev:2006ct}. As shown in \cite{Andreev:2006ct}, this
potential has the same behavior expected from QCD, ie it is linearly
rising at large distances while presenting a Coulomb behavior at
small distances. Clearly the interpolation between these two
behaviors is phenomenologically relevant if it corresponds to length
scales typical of the hadrons. This is a possible way to
discriminate among the different models. Another aim of this paper
is therefore to find the region where the AdS inspired potential
differs from the usual QCD-based potentials and to see if it is of
phenomenological significance.  We fix the parameters of the model
using, as an input, data from the meson spectrum. We compare our
results with the available experimental data for the mesonic
spectrum and we find a reasonable good agreement. Once the
parameters of the model are fixed,  we can  compute the diquark
masses, ie a set of phenomenological parameters that we use,
together with some additional hypothesis, to predict other spectra.

The results of the paper are as follows. Assuming a
diquark-antidiquark structure of tetraquark states, we find masses
for some of the $X$ and $Y$ states in reasonable agreement with
experiment. The same procedure can be applied to tetraquarks
comprising  one charm quark and hidden strangeness. In this case the
results are hardly compatible with the masses of possible tetraquark
candidates, ie the states
 $D_s(2317)$
\cite{Aubert:2003fg}, $D_s(2457)$ \cite{Aubert:2003fg} and $X(2632)$
\cite{Evdokimov:2004iy}.

The paper is organized as follows. In section \ref{II} we describe
the model, introducing the Salpeter equation, the potential term and
the numerical method adopted to solve the wave equation. In section
\ref{III} we determine the parameters of our model fitting meson
spectra; the results allow a comparison with the presently available
meson data. In section \ref{IV} the constituent diquark masses are
calculated, in the one-gluon exchange approximation. In section
\ref{VI} we compute the spectra of a few tetraquark states. Finally,
in section \ref{VII} we draw our conclusions.

\section{\label{II}Wave equation and the central potential}
The wave equation we will use in this paper is an eigenvalue
equation for the hamiltonian of two point-like particles, taking
into account the reltivistic kinematics. It is known in the
literature as Salpeter equation. In QCD it arises from the
Bethe-Salpeter equation replacing the interaction with an
instantaneous local potential $V(r)$ and considering a limited Fock
space containing $q\bar q$ pairs only.  We write the Salpeter
equation for a meson comprising a quark and an antiquark in the
meson rest frame as follows ($\hbar=c=1$):
\begin{equation}\label{salpeter}
\left(\sqrt{m_1^2-\nabla^2}+\sqrt{m_2^2-\nabla^2}+V(r)\right)
\psi({\bf r})\,=\,M\, \psi({\bf r})\ .
\end{equation}For central potential one can search for energy eigenfunctions
with definite angular momentum $\ell$ thus writing \be \psi({\bf
r})\,=\, Y_{\ell\,m}({\bf\hat r})\phi_\ell(r)\ . \ee The radial
wavefuction $\phi_\ell(r)$ satisfies the equation \be \frac{2}{\pi}
\,\int_0^{+\infty}dr^\prime\,r^{\prime\,2} \,\int_0^{+\infty} dk
\left(\sqrt{k^2+m_1^2}\,+\,\sqrt{k^2+m_2^2}\right)\,k^2\,
j_\ell(kr)\,
j_\ell(kr^\prime)\phi_\ell(r^{\prime})\,=\,[M-V(r)]\phi_\ell(r)~.
\label{salpeter3d4} \ee where $j_\ell(x)$ are spherical Bessel
functions.  For $\ell=0$ $j_0(x)=\sin x/x$ and the Salpeter equation
reduces to \be \frac{2}{\pi} \,\int_0^\infty dr^\prime\,
\,\int_0^\infty
dk\,\left(\sqrt{k^2+m_1^2}\,+\,\sqrt{k^2+m_2^2}\right)
\,\sin\left(kr\right)\, \sin\left(kr^\prime\right)u_0(r^{\prime})\,
=\,[M-V(r)]u_0(r)\label{salpeter3d5}\,, \ee where \be
u_0(r)=r\phi_0(r)\,.\ee Let us discuss the potential energy. The
potential $V(r)$ we adopt comprises two parts: \be\label{pot2}
V(r)=V_{AdS}(r)+V_{spin}(r)\ .\ee $V_{AdS}(r)$ describes the central
term of the potential, ie it contains  its linearly confining part
at large distances and the short distance behavior predicted by
perturbative QCD. For $V_{AdS}(r)$ we use a slightly modified
version of the potential obtained by Andreev and Zakharov
\cite{Andreev:2006ct} in the framework of the gauge/string duality
approach \cite{ads/cft,malda}. $V_{spin}(r)$ is the hyperfine term
to be discussed below.
\subsection{Static quark potential} In the gauge/string duality
approach \cite{ads/cft,malda} the expectation value of the Wilson
loop is given by
\begin{equation}\label{wloop}
\langle\,W({\cal C})\,\rangle \sim e^{-S} \,,
\end{equation}
where $S$ is an area of a string world-sheet bounded by a curve
$\cal C$ at the boundary of the AdS space (for references to the
original papers see also \cite{ADSCFT}). To compute the potential
the authors start with the Nambu-Goto action
\begin{equation}\label{action}
S=\frac{1}{2\pi\alpha'}\int d^2\xi\,\sqrt{\det \,
G_{nm}\,\partial_\alpha X^n\partial_\beta X^m} \,.
\end{equation}
with the following background metric in $D=5$
\begin{equation} \label{metric}
ds^2=G_{nm} dX^n dX^m= R^2\frac{h}{z^2} \left(dx^i dx^i+dz^2\right)
\,,\quad\quad h=\exp{\frac{cz^2}2}\,,
\end{equation}
where $i=0,\dots ,3$.  Choosing in (\ref{action}) $\xi^1=t$ and
$\xi^2=x$, Eq.  (\ref{action}) becomes
\begin{equation}\label{action2}
S=\frac{g }{2\pi}T\int^{+\frac{r}{2}}_{-\frac{r}{2}}
dx\,\frac{h}{z^2} \sqrt{1+\left(\frac{d z}{dx}\right)^2} \,,
\end{equation}
where $\displaystyle g=\frac{R^2}{\alpha'}$. From this action one
can obtain an equation of motion for the variable $z$, from which an
expression for the interquark distance $r$ is derived as follows
\cite{Andreev:2006ct}:
\begin{equation}
r(\lambda)\,=\,2\, \sqrt{\frac{\lambda}{c}} \int_0^1 dv v^{2}
\exp^{\lambda (1-v^2)/2} \left(1-v^4
\exp^{\lambda(1-v^2)}\right)^{-1/2}\label{r}\ .
\end{equation} This expression depends on the constant $c$;
$\lambda$ is a parameter in the range $]0,\,2[$. In terms of the
original parameters it is given by $\lambda=c z_0^2$, with $z_0$ the
value of the fifth coordinate $z$ in $x=0$.

The potential is obtained by computing the energy of the
configuration; first one changes the integration variable from $x$
to $z$ in (\ref{action2}); the resulting integral is divergent at
$z=0$ and has to be regularized. The finite part gives
\begin{equation}\label{zakpot} E(\lambda)\,=\,\frac{g}{\pi}
\sqrt{\frac{c}{\lambda}} \left( -1+\int_0^1 dv v^{-2} \left[
\exp^{\lambda v^2/2} \left(1-v^4
\exp^{\lambda(1-v^2)}\right)^{-1/2}-1\right]\right)\ .
\end{equation}This is the potential given in \cite{Andreev:2006ct}.
The dependence $E(r)$ is obtained by elimination of the parameter
$\lambda$ between (\ref{r}) and (\ref{zakpot}); $E(r)$ depends on
two parameters $g$ and $c$.

Our potential $V_{AdS}(r)$ is given as follows:\be
V_{AdS}(r)\,=\left\{
\begin{array}{cc}
\label{pot}
E(r)\,+\,V_0 & \hskip1.2cm r>r_m\\
E(r_m)\,+\,V_0 & \hskip1.8cm r\le r_m \hskip.5cm .
\end{array}
\right. \label{adspotential}\ee

This phenomenological procedure, which has allowed to get an
expression for the QCD potential with the expected properties, is
known as the bottom-up approach of AdS/QCD correspondence. According
to it, starting from QCD, one constructs its dual theory in such a
way to reproduce some properties of QCD. Besides the QCD potential,
many results have been obtained so far, such as the determination of
the numerical values of some observables \cite{Erlich:2005qh}, in
good agreement with experimental data, or the linearity of the Regge
trajectories \cite{Karch:2006pv,Andreev:2006vy} .

The expression (\ref{zakpot}) derives from the metric
(\ref{metric}); we have also tried to use a slight modification of
the dilaton term in (\ref{metric}), specifically, changing the power
of $z$ at the exponent. We find that, for any power of $z$, the
potential has the same behavior, ie Coulomb-like for small $r$ and
linearly rising for large $r$. As far as the QCD potential is
concerned  they are all possible candidates for the AdS metric of
the QCD dual space.

 We have introduced a constant term $V_0$ in the potential.
 This is allowed because the more general
 string action is obtained by adding to the Nabu-Goto action a
 term\begin{equation}\label{action2b}
\delta S=C_R\int d^2\xi\,R \,\sqrt{\det \, G_{nm}\,\partial_\alpha
X^n\partial_\beta X^m}\,,
\end{equation}where $R$ is the scalar curvature. One can easily prove
that adding $\delta S$ to $S$ corresponds to add a constant term to
the interquark potential.

Another modification appearing in (\ref{pot}) is the introduction of
a  cutoff at short distances $r\le r_m$ \cite{Colangelo:1990rv}. The
reason lies in the use of the relativistic kinematics embodied in
Eq. (\ref{salpeter}). As a matter of fact, the potential
$E(\lambda)$ in (\ref{zakpot}) diverges as $\lambda\to 0$. It is a
Coulombic divergence occurring at $r\to 0$, because the small
distance region corresponds to $\lambda\to\,0$. This divergence is
harmless if one uses the Schr\"odinger equation for the
wavefunction, but in the case of the Salpeter equation
(\ref{salpeter}) it produces an unphysical logarithmic divergence
for the $s-$wave wavefunction for $r\to 0$. To cure this unphysical
singularity arising from the static approximation one assumes a
constant potential for $r$ smaller than\be r_m\sim\frac{1}{M}\,.\ee
The proportionality constant can be fixed using quark duality
arguments
 \cite{Cea:1986bj}, which shows that one expects different results for
  the equal and unequal mass cases. Thus we use
  \bea
r_m&=&\frac{k}{M} \hskip2cm (m_1=m_2)\,,\cr&&\cr
r_m&=&\frac{k^\prime}{M}\hskip2cm(m_1\neq m_2)\,. \eea This
concludes the discussion of the central part of the potential.
\subsection{Spin term}
 Let us now discuss the spin term.
In the  approximation of one-gluon exchange one knows that the spin
term is enhanced at short distances and is proportional to inverse
quark masses.  Following \cite{Barnes:2005pb} we use
\begin{equation}\label{spin}
V_{spin}(r)\,=\,A \frac{\tilde\delta(r)}{m_1 m_2}{\bf S_1}\cdot{\bf
S_2}
\end{equation}where $\tilde\delta$ is a function enhanced at
small distances. 
In the constituent quark model the constant $A$ is proportional to $\alpha_s$, 
which is a running coupling constant, so we have introduced two different 
parameters for mesons containing at least one charm and one bottom quark 
($A_c$ and $A_b$, respectively).
We adopt the
smeared delta function used in ref. \cite{Barnes:2005pb}:\be
\tilde\delta(r)=\left(\frac{\sigma}{\sqrt{\pi}}\right)^3
e^{-\sigma^2 r^2}\ .\ee The potential depends therefore on
8 parameters: $c,\,g,\,V_0,\,k,\,k^\prime,\,A_c,\,A_b,\,\sigma$. Moreover we
have to fix the constituent quark masses: $m_u=m_d\equiv m_q$, $m_s$,
$m_c$, $m_b$.
\subsection{Numerical solutions}
To solve Eq. (\ref{salpeter}) we use the Multhopp method
\cite{Karamcheti}, which, as shown in \cite{Colangelo:1990rv}, is
particularly useful for equations of the form considered here. By
this method one transforms the integral equation in a set of linear
equations introducing $N$ parameters $\theta_k$, called Multhopp's
angles. The set of equations is as follows (we refer the reader to
\cite{Colangelo:1990rv} for further details):
\be\sum_{m=1}^{N}B_{km}\psi(\theta_k)\,=\,M\psi(\theta_m)\ee where $
\psi(\theta_k)= u_0(-\cot\theta_k)$\,,
$\theta_k=\frac{k\pi}{N+1}\,\hskip1cm(k=1,\cdots N)$ and
 \be
B_{km}\ =\ \frac{2}{N+1}\sum_{j=1}^N\sin(j\theta_m)I_{jk}\ee with
 \bea &&I_{jk}=\,-\,\frac2\pi\lim_{\epsilon\to
 0}\left\{\int_0^{\theta_k-\epsilon}\frac{d\theta\sin j\theta}{\sin^2\theta
 (\cot\theta-\cot\theta_k)^2}+\int^\pi_{\theta_k+\epsilon}
 \frac{d\theta\sin j\theta}{\sin^2\theta(\cot\theta-\cot\theta_k)^2}
 \,-\,\frac2\epsilon\sin^2\theta_k\sin j\theta_k\right\}\cr&&+
\frac2\pi\,\int_0^{\pi}\frac{d\theta\sin j\theta}{\sin^2\theta}
\Big[\frac1{
(\cot\theta-\cot\theta_k)^2}\,-\,\frac1{2|\cot\theta-\cot\theta_k|}
\Big( m_1K_1(m_1|\cot\theta-\cot\theta_k|)+(m_1\leftrightarrow m_2)
\Big)\Big].~~~~
 \label{k1}\eea
 Notice that the writing in (\ref{k1}) makes explicit the prescription
 to avoid the divergence in the integral both
 for the massive and the massless case; $K_1(x)$ is a
  modified Bessel function.

\section{\label{III}Meson spectrum}

The strategy we follow to evaluate the parameters of the model is to
estimate them by using information from the meson spectra.
Subsequently, in the next sections we use this information to fit
diquark masses and, from these data, tetraquark masses. In principle
one might follow a different strategy, using all available
experimental data to get a best fit of the parameters. We have not
followed this approach  because to get the spectrum of diquarks  and
tetraquarks we make further assumptions, e.g. the use of the
quark-quark potential as obtained by one-gluon-exchange
approximation.

\renewcommand{\arraystretch}{.9}\begin{table}[ht!]
\caption{\label{tab:spettriheavy}Mass spectra for heavy  mesons;
$q=u,\,d$. A star $(*)$ means that this state needs confirmation.
Units are GeV.}
\begin{center}
\begin{tabular}{|c|c||c|c|c|c|c|c|}
\hline
Flavor & Level &  \multicolumn{3}{|c|}{$J=0$} & \multicolumn{3}{|c|}{$J=1$} \\
\cline{3-8}
 & & Particle & Th. mass& Exp. mass  \cite{PDG} & Particle & Th. mass & Exp. mass \cite{PDG} \\
 \hline
$c\bar q$ & $1S$ &  $D$ & $1.862$ &$1.867 $& $D^*$ &2.027
 &$2.008$  \\
& $2S$  & &3.393 &  &   &2.598  &$2.622$  \\
& $3S$  & &   2.837 & &   &  2.987&  \\
\hline
$c\bar s$ & $1S$  & $D_s$ & 1.973&$1.968$ &  $D_s^*$  &2.111 & $2.112$ \\
& $2S$  &&  2.524&  & &2.670 &  \\
& $3S$  &  &2.958&  & &3.064&  \\
\hline $c\bar c$  & $1S$  & $\eta_c$  &  2.990&$2.980$  & $J/\psi$
& 3.125&$3.097$  \\
& $2S$  & &  3.591 &$3.637$ &  &  3.655& $3.686$  \\
& $3S$  &   &3.994 & & &   4.047& $4.039$  \\
\hline
$b\bar q$    & $1S$  & $B$   &5.198 & $5.279$  &   $B^*$     & 5.288&$5.325$  \\
& $2S$  && 5.757&  &  & 5.819&  \\
& $3S$  & &  6.176&  &&  6.220&  \\
\hline
$s\bar b$    & $1S$   & $B_s$ &5.301& $5.366$  & $B_s^*$&5.364&$5.412$  \\
& $2S$  & &  5.856 &  &  &  5.896&  \\
& $3S$  && 6.266&  & & 6.296&  \\
\hline
$b\bar c$ & $1S$ & $B_c$ &6.310& $6.286 $&$B_c^*$&  6.338 &$6.420$ \\
& $2S$  & & 6.869&  &  &   6.879 & \\
& $3S$  & & 7.221 &  &   &      7.228 &  \\
\hline $b\bar b$& $1S$& $\eta_b$&9.387&$9.300^{*}$&$
\Upsilon$ & 9.405&$9.460$  \\
& $2S$  && 10.036& && 10.040&$10.023$ \\
& $3S$  &   & 10.369& &  &   10.371&$10.355$\\
& $4S$  &  & 10.619   & & &    10.620&$10.579$ \\
\hline
\end{tabular}
\end{center}
\end{table}

 Therefore, after having fixed the parameters from the
mesonic spectrum, we will use below experimental data for $X$ and
$Y$ states to test the hypothesis of diquark-antidiquark structure
of these exotica.

 In order to get the numerical values for the parameters
($c,\,g,\,V_0,\,k,\,k^\prime,\,A_c,\,A_b,\,\sigma$, $m_u=m_d\equiv m_q$,
$m_s$, $m_c$, $m_b$) we use the Salpeter equation for mesons with
both $J^P=0^-$ and $1^-$. We consider only mesons containing at
least one heavy ($c,\,b$) quark, because we expect that the
approximation of static potential works better for these states.The
set of available data comprises about 20 masses. The results of the best fit
 are reported in Table 
\ref{tab:spettriheavy} for heavy mesons. They correspond to the
following set of parameters:  \bea &&c\,=\,0.3 \mbox{ GeV}^2\,, ~~~~~~
g\,=\,2.75\,, ~~~~~~
 ~~~ V_0\,=\,-0.49 \mbox{ GeV}\cr
 && A_c=7.92 \mbox{ GeV}^3 \,, ~~~ A_b=3.09 \mbox{ GeV}^3\cr
&&k\,=\,1.48\,, ~~~~~~ k^\prime\,=\,2.15\,, ~~~~~~\sigma=1.21 \mbox{ GeV}^{2}\cr&& m_q=0.302
\mbox{ GeV} \,, ~~~ m_s=0.454 \mbox{ GeV} \,, ~~~ m_c=1.733 \mbox{
GeV} \,, ~~~ m_b=5.139 \mbox{ GeV}\,.\label{parameters}\eea

In order to test the limits of the model we also compute the masses
of states containing only light ($u,\,d,\,s$) quarks. 
The results for their spin averaged masses are shown in Table \ref{tab:spettrilightbar}. 
For the lightest mesons we find a large deviation, while such a discrepancy
is somehow reduced in case of $s\bar{s}$ and for higher radial excitations. This is 
reasonable, since a constituent quark model with instantaneous interaction 
is not able to describe the chiral dynamics of light states. 
The better accuracy of the $s\bar{s}$ system allows us to fix the spin constant 
 $A_s$ in eq. (\ref{spin}) from this channel, obtaining $A_s=11.3$ GeV$^3$, which 
 will be used in the calculation of the masses of the lightest diquarks. 
 With this value, we obtain for $\varphi$ a mass m=1.011 GeV (the experimental 
 value is 1.019 GeV) and for  
 $\varphi^\prime$ a mass m=1.663 GeV (the experimental value is 1.680 GeV).

\renewcommand{\arraystretch}{.9}\begin{table}[h]
\caption{\label{tab:spettrilightbar}Mass spectra for spin averaged masses of light  mesons;
$q=u,\,d$. Units are GeV.} \begin{center}
\begin{tabular}{|c|c|c|c|}
\hline Flavor & Level & Th. mass &  Exp. mass \cite{PDG}  \\
\hline
 $q\bar q$ & $1S$ & 0.792  &$0.616$ \\
& $2S$  & 1.386 &   $1.424$  \\
\hline $q\bar s$&$1S$& 0.932&$0.794$   \\
& $2S$  & 1.501& $$   \\
\hline $s\bar s$ & $1S$& 0.981 &$0.912$\\
& $2S$  &  1.571   &  $\approx 1.653$  \\
\hline\end{tabular}
\end{center}
\end{table}

 It might be useful at this stage
to study the effect of the relativistic kinematics on the equation
of state. To this end we have used the same potential, with the
values of parameters indicated by Eq. (\ref{parameters}) with two
different equations: the Salpeter equation (\ref{salpeter}) and the
Schr\"odinger equation. The results of the two equations for mesons
with $J^P=1^-$ are reported in Table \ref{comparison}.  The
comparison between the two computed spectra and the experimental one shows that, as expected,
 the results obtained by the Salpeter equation are more accurate then
 the ones obtained by the Schr\"odinger equation.  
 We conclude that the advantage of using the Salpeter
equation is particularly significant for the charmed states, since
this equation takes into account a relevant source of corrections,
i.e. those due to the relativistic kinematics.
\renewcommand{\arraystretch}{.9}
\begin{table}[ht!]\caption{\label{comparison}
Comparison of spectra for  mesons with $J^P=1^-$ computed by the
Salpeter equation in Eq. (\ref{salpeter}) and the Schr\"odinger
equation. The same potential $V(r)$ is used in both cases.}
 \begin{center}
\begin{tabular}{|c|c|c|c|c|}
\hline Flavor & Level &  Salpeter &Schr\"odinger   \\
\hline
$c\bar q$ & $1S$  & 2.027 & 2.154 \\
& $2S$ &    2.598&2.877   \\
\hline
$c\bar s$ & $1S$&2.111 &2.182  \\
& $2S$   &  2.670 & 2.843 \\
\hline $c\bar c$  & $1S$ &
 3.125 &3.133  \\
& $2S$  &   3.655&3.695   \\
\hline
$b\bar q$    & $1S$&    5.288&5.494  \\
& $2S$ &   5.819& 6.204 \\
\hline
$s\bar b$    & $1S$  & 5.364&5.507  \\
& $2S$ &   5.896&6.154  \\
\hline
$b\bar c$ & $1S$ &  6.338 &6.550 \\
& $2S$  & 6.879 &6.922 \\
\hline
$b\bar b$& $1S$ & 9.405&9.774  \\
& $2S$  & 10.040&10.055 \\
\hline
\end{tabular}
\end{center}
\end{table}

A final comment is for the use of the AdS/QCD inspired potential. As
we stressed in the introduction, many potentials have been used in
the literature to fit meson spectra. If they have to reproduce the
constraints of QCD they must rise linearly at large distances and
have a Coulomb-like behavior at small $r$. We wish to compare the
potential we have used, i.e. $V_{AdS}=E(r)+V_0$ in Eq.
(\ref{adspotential}), with a typical QCD-inspired potential, i.e.
the Richardson's potential \cite{Richardson:1978bt}: \be
V_R(r)=\frac{8\pi}{33-2 n_f}\Lambda\left(\Lambda r-\frac{f(\Lambda
r)}{\Lambda r}\right)\,+V_1\,,\label{rich}\ee where $\Lambda$ is a
parameter, $n_f=3$ is the number of flavors and \be
f(t)=\frac{4}{\pi}\int_0^{\infty}dq\,\frac{\sin
(qt)}{q}\left(\frac1{\ln(1+q^2)}\,-\,\frac1{q^2}\right)\,.\ee

\begin{figure}[ht]
\begin{center}
\includegraphics[width=9cm]{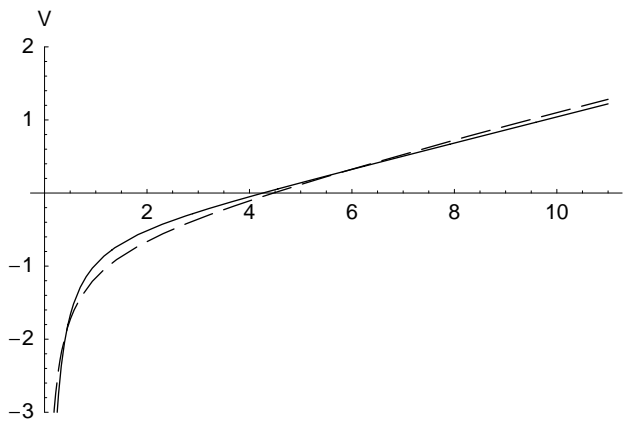}~~~~~~~~~~~~~~~~~~~~~~~~~~~
\end{center}
\caption{\label{Fig:Fig0} Comparison between  the AdS/QCD inspired
potential $V_{AdS}(r)$ (solid line) and the Richardson's potential
\cite{Richardson:1978bt} $V_R(r)$ (dotted line). For the values of
the parameters see the text.}
\end{figure}

To allow for a comparison, similarly to what we have done for the
AdS potential we have added a constant term $V_1$ to the original
Richardson's potential. We fix the value of $\Lambda$ and $V_1$
imposing that the linearly confining parts of $V_{AdS}$ and $V_R$
coincide at large $r$.  For   $V_{AdS}$ we use the fit obtained by
the meson spectrum; this gives for $\Lambda$ and $V_1$ the values
$\Lambda=0.44$ GeV and $V_1\sim\,-\,0.6$ GeV.

We are now able to compare the two potentials and the result is in
Fig. \ref{Fig:Fig0}. It shows that, even though the two potentials
almost coincide asymptotically, they differ significantly
in the intermediate region, which is the region of interest from a
phenomenological point of view.

\section{\label{IV}Diquark masses}
In view of the applications to be discussed in the subsequent
sections we wish to derive a set of parameters that we call {\it
effective diquark masses} or simply diquark masses.

In the one-gluon exchange approximation the color antisymmetric
channel corresponding to antitriplet in the decomposition\be 3
\otimes 3= \bar{3}\oplus6\ee is attractive. Moreover, in the same
approximation the attractive potential between the two quarks $QQ$
is half the one between a quark $Q$ and an antiquark $\bar Q$. We
therefore adopt, as usual, the following value for the potential
between the two quarks comprising the diquark: \be
V_{QQ}=\frac12V_{Q\bar Q}\,.\label{QQ}\ee However it must be kept in
mind that this approximation holds only for the QCD perturbative
interaction and might be modified in the linearly confining part.
Therefore we expect it works better for heavy diquarks, as, in this
case, the interaction explores smaller distances than for light
diquarks.

To derive the diquark effective masses we use the Salpeter equation
with the same set of parameters used to fit the meson spectrum .
Also in this case, as for mesons, one has that the spin term works
differently for $S=1$ and $S=0$ states. It adds a positive
contribution for $S=1$ and subtracts the same contribution
multiplied by a factor of $-\,3$ for $S=0$.

\renewcommand{\arraystretch}{.9}
\begin{table}[ht!]
\caption{\small Diquark masses\label{diquark}. The masses of the
present paper are
 obtained by the
Salpeter equation (see text).  The model in
\cite{Ebert:2005nc,Ebert:2007rn} uses a quasipotential of the
Schr\"odinger type \cite{Ebert:2002pp}.$\{QQ\}$ (resp. $[QQ]$) means
a spin 1 (resp. $S=0$) diquark $QQ$.  Units are GeV.
\label{4qcqcq}}\begin{center}
\begin{tabular}{|c|c|c||c|c|c|}\hline
 State& Mass (this paper)&Mass \cite{Ebert:2005nc,Ebert:2007rn}&Mass&
  Mass (this paper)&Mass\cite{Ebert:2005nc,Ebert:2007rn}\\
\hline \{qs\}& 0.980&1.069&[qs]&0.979&0.948\\
\hline \{ss\}& 1.096&1.203&&&\\
\hline \{cq\}& 2.168&2.036&[cq]&2.120&1.973\\
\hline \{cs\}& 2.276&2.158&[cs]& 2.235&2.091\\
\hline \{cc\}& 3.414&3.226&&& \\
\hline \{bq\}& 5.526&5.381&[bq]&5.513&5.359\\
\hline \{bs\}& 5.630&5.482&[bs]&5.619&5.462\\
\hline \{bc\}& 6.741&6.526&[bc]& 6.735&6.519\\
\hline \{bb\}& 10.018&9.778&&& \\
\hline
\end{tabular}
\end{center}
\end{table}

The results are reported in Table \ref{diquark}; diquarks with $S=0$
are denoted as $[QQ]$, those with $S=1$ as $\{QQ\}$. We have also
reported the results of a different fit found by the authors of
Refs. \cite{Ebert:2005nc,Ebert:2007rn} who use a quasipotential of
the Schr\"odinger type \cite{Ebert:2002pp}. We also note that in a
series of papers
\cite{Maiani:2004uc,Maiani:2004vq,Maiani:2005pe,Maiani:2007vr}
Maiani and collaborators have put forward an interpretation of the
new $X$ and $Y$ states, as well as of light scalars such as
$\sigma(480)$, $\kappa(800)$, $f_0(980)$, etc., as four-quark states
comprising a diquark and an antidiquark (see below). They use a
quark constituent model similar to the well known De
Rujula-Georgi-Glashow model \cite{De Rujula:1975ge}. Also in
\cite{Maiani:2004uc,Maiani:2004vq,Maiani:2005pe,Maiani:2007vr} an
effective diquark mass is used, however its meaning is different
from ours, because the constituent quark mass used in these papers
takes into account both the average kinetic energy and the potential
energy, differently from our case, where a wave equation is
considered. Therefore no numerical comparison is possible among our
values for diquark masses and those found in
\cite{Maiani:2004uc,Maiani:2004vq,Maiani:2005pe,Maiani:2007vr}.
\section{\label{VI}Tetraquark spectrum}
In this section we wish to discuss the possibility that two
diquarks, more precisely a diquark and an antidiquark, combine to
produce a tetraquark state. Such an interpretation was advanced long
ago in order to give an interpretation of light scalars $a_0(980)$
and $f_0(980)$ in terms of constituent quarks
  \cite{Jaffeold}. The
recent discovery of new states, with both hidden and open charm,
 has raised
a new interest for this model, though the interpretation of the new
states is
 controversial, see
 for reviews \cite{Swanson:2006st} and \cite{Jaffe:2004ph}.

Let us start with the state $X(3872)$
\cite{Choi:2003ue,Abazov:2004kp,Acosta:2003zx,Aubert:2004ns,Abe:2005ix,Abe:2005iya}.
The average mass of this state  is $3871.2~\pm~0.4$ MeV and its
quantum numbers should be $J^{PC}=~1^{++}$. The  assignment $C=+1$
follows from the fact that the decay $X\to \gamma J/\Psi$ is
observed.
 Moreover the decay $X\to \pi^+\pi^- J/\Psi$ is also
observed; the part of the $2\pi$ invariant mass spectrum that can be
ascribed to a $\rho^0$ decay is
 consistent with
$S-$wave decay of the $X$ state. From this the assignment $P=+1$
follows. Finally the angular distribution in this channel is
incompatible with $J=0$ and therefore the only remaining possibility
is $J=1$ or $J=2$. If the peak in the $D^0\bar{D}^0\pi^0$ decay
channel at 3875.4 MeV (at $2\sigma$ from the mass of $X(3872)$) is
interpreted as due to this state, then the $J=2$ should be excluded,
which leaves us with $J=1$ only.

Several interpretations have been proposed for this state. Since the
mass of this state almost coincides with the sum of the masses of
the $D^0$ and $\bar D^{*0}$ states, a natural explanation is that of
a molecular state comprising the two charmed mesons
\cite{Tornqvist:2004qy,Swanson:2003tb,Close:2003sg,Pakvasa:2003ea,Wong:2003xk,Braaten:2004rw}.
The interpretation as a $\chi_{c1}^\prime$ state
\cite{Barnes:2003vb} is
 unlikely because of the small value of
 the ratio $\dd\frac{{\cal B}(X\to \gamma
J/\Psi)}{{\cal B}(X\to \pi^+\pi^- J/\Psi)}$ and for the value of the
mass. Also the interpretation as a $c\bar cg$ hybrid
\cite{Li:2004sta} is difficult due to the fact that lattice data
predict larger masses for these states (for a discussion and for a
comprehensive list of other models see \cite{Swanson:2006st}).

 As mentioned above, another possibility is
that this state comprises four quarks. They might be a four quark
cluster without internal structure \cite{Hogaasen:2005jv} or a bound
state of two diquarks, as first discussed in \cite{Maiani:2004vq}
and more recently in \cite{Ebert:2007rn}. We shall return to this
interpretation below.

 The Belle
collaboration observes two bumps,  at 3940 MeV. They do not
necessarily correspond to two different states. The state called
$Y(3940)$ is
 observed in the decay mode $B~\to~ K\omega\, J/\Psi$
\cite{Abe:2004zs}. Its reported mass is $M=3943\pm11\pm13$
 MeV. Its interpretation as a
charmonium $c\bar c$ state is possible, but should be corroborated
by the observation of the decay mode $Y\to D^{(*)}\bar D^{(*)}$,
which has not yet been seen. Also the interpretation as a $c\bar
c$-gluon has been proposed, though
 the predicted mass of such a state, around 4.3-4.5
GeV from lattice QCD computations \cite{Bernard:1997ib,Mei:2002ip},
is significantly larger than the measured value. Also in this case a
four quark interpretation is possible, in particular the state might
be the $2^{++}$ state predicted by the diquark-antidiquark scheme
\cite{Maiani:2004vq}. The other state $X(3940)$ needs confirmation;
it is observed \cite{Abe:2005hd} in  double charm production:
$e^+e^-\to J/\psi X\to J/\psi D\bar D^*$. Its possible
interpretations are the states $\chi^\prime_c$ or
$\eta_c^{\prime\prime}$ \cite{Swanson:2006st}.

 We wish to test the interpretation of these states
 as diquark-antidiquark bound states, in the same
 spirit of Refs.
\cite{Maiani:2004uc,Maiani:2004vq,Maiani:2005pe,Maiani:2007vr} and
\cite{Ebert:2005nc}. Note however that our approach is different
from that followed in these papers. As a matter of fact we apply a
wave equation for the tetraquark states, comprising the diquarks
$(Q_1Q_2)$ and $(\bar Q_3\bar Q_4)$, as follows
\begin{equation}\label{salpeterbis}
\left(\sqrt{m_{12}^2-\nabla^2}+\sqrt{m_{34}^2-\nabla^2}+\tilde
V(R)\right) \psi_t({\bf R})\,=\,M_t\, \psi_t({\bf R})\ .
\end{equation}
Here $M_t$ and $\psi_t({\bf R})$ are  the tetraquark mass and
wavefunction respectively, $m_{ij}$ is the effective diquark mass
computed above; $R$ is the distance between the centers of the two
diquarks. We take into account the structure of the diquarks by
defining a smeared potential as follows:\be \tilde V(R)=\frac1N\int
d{\bf r_1}\int d{\bf r_2}|\psi({\bf r_1})|^2|\psi({\bf r_2})|^2
V\Big(\Big|{\bf R}+{\bf r_1}-{\bf r_2}\Big|\Big)\dd
\label{n1}\ee with\be N=\int d{\bf r_1}\int d{\bf
r_2}|\psi_{12}({\bf r_1})|^2|\psi_{34}({\bf r_2})|^2\ .
\label{n2}\ee In these equations $\psi_{12}({\bf r_1}) $ (resp.
$\psi_{34}({\bf r_2}) $) is the wave function of the diquark whose
content is $Q_1Q_2$ (resp. $Q_3Q_4$); for $V(r)$ we use
(\ref{pot2}), and the masses appearing in the spin term of Eq.
(\ref{spin}) are diquark masses.
 In this paper we only consider diquark with internal orbital quantum
  number $\ell=0$; therefore
 $\psi_{ij}$ is related to $u_0(r)$ defined above as follows:
 $\dd\psi_{ij}({\bf r})= \frac{u_0(r)}{r}$.
 
\begin{figure}[ht]
\begin{center}
\includegraphics[width=9cm]{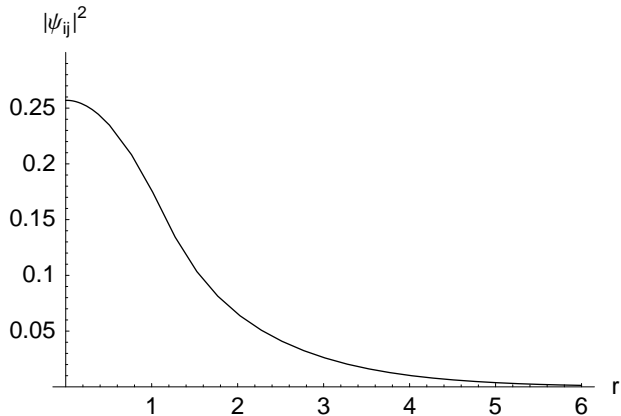}~~~~~~~~~~~~~~~~~~~~~~~~~~~
\end{center}
\caption{\label{Fig:Fig1} The squared diquark wavefunction (in
GeV$^3$) versus $r$ (in GeV$^{-1})$. Data correspond to the diquark
$[cq]$.}
\end{figure}

\begin{figure}[ht]
\begin{center}
\includegraphics[width=9cm]{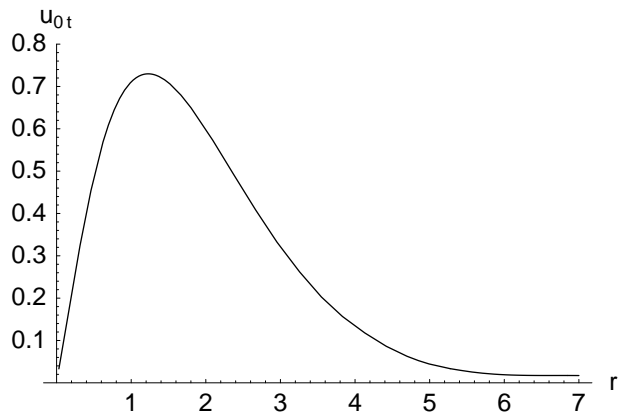}~~~~~~~~~~~~~~~~~~~~~~~~~~~
\end{center}
\caption{\label{wft} The tetraquark wavefunction $u_{0t}(r)$ (in
GeV$^{1/2}$) versus $r$ (in GeV$^{-1})$. Data correspond to the tetraquark
$[cq]\{\bar{c}\bar{q}\}$.}
\end{figure}

Since $|\psi_{12}({\bf r})|^2$ is strongly peaked at
$r\sim 0$ (see Fig. \ref{Fig:Fig1}) we cut-off the integrals in
Eqns. (\ref{n1}) and (\ref{n2})  at $\displaystyle r_{1,2}\le R_0$
where $R_0$ is the peak value of
 $u_{0t}(r)$, the wavefunction of the tetraquark.
 This procedure ensures that the two diquarks are on
 average inside the tetraquark's bag. In Fig. \ref{wft} is represented 
 $u_{0t}(r)$ for the tetraquark $[cq]\{\bar{c}\bar{q}\}$.
 
\begin{figure}[ht]
\begin{center}
\includegraphics[width=9cm]{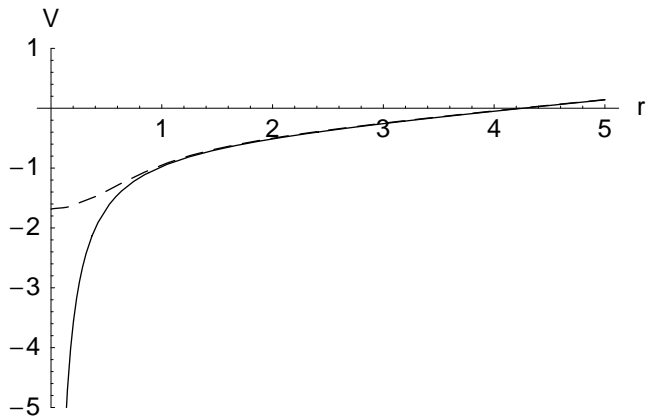}~~~~~~~~~~~~~~~~~~~~~~~~~~~
\end{center}
\caption{\label{Fig:Fig2} The potential between static diquarks
(dashed line) and its modification according Eq. (\ref{n1}) (solid
line). Units are. GeV (V) and GeV$^{-1}$ (r). Data refer to the
$[cq]\{\bar c\bar q\}$ potential.}
\end{figure}

 We finally note that we use for $\psi_{12}({\bf r})$
the result obtained from the
 diquark wave equation. This is only approximately correct
  because that equation provides the diquark wavefunction in the diquark rest frame,
  whereas (\ref{salpeterbis}) holds in the tetraquark rest frame. However
  for diquarks comprising heavy quarks ($c,\,b$) the average diquark velocity is small
  (we estimate $\beta\sim 0.15$ for diquarks with open charm and  $\beta\sim 0.06$ for
  diquarks with open bottom). Therefore we can neglect the distortion induced by the Lorentz boost on the wavefunction.
The effect of convoluting the potential with the wavefunctions in
(\ref{n1}) is shown in Fig. \ref{Fig:Fig2}.

In Tables \ref{T1} and \ref{T2} we present our predictions for the
four-quark states with hidden charm  and hidden bottom respectively.
Notice that we have adopted the same values of the free parameters
used in the previous sections, so basically our calculation is
parameter-free. It is remarkable that in the two cases when a
comparison with experiment is possible, our results agree, within
the errors, with the data. A further comment is the following. A
peculiar feature of the diquark-antidiquark scheme for the $X$ state
is the prediction of four different states with mass differences of
a few MeV. Two of them are neutral: $X_u=[cu][\bar c\bar u]$,
$X_d=[cd][\bar c\bar d]$, and two charged: $X^+=[cu][\bar c\bar d]$,
$X^-=[cd][\bar c\bar u]$. It is possible that the two states seen
through the decay modes $J/\psi\pi^+\pi^-$ and $D^0\bar D^0\pi^0$
are really different and they would correspond to the neutral states
$X_u$, $X_d$ since the mass difference between the two bumps  at
3871 and 3876 MeV is of the same order of magnitude of the mass
difference between the $u$ and $d$ quarks. According to
\cite{Maiani:2007vr} this is a piece of evidence in favor of the
four quark interpretation of these states. We have not included any
mass  difference between $u$ and $d$ quarks, but also in our scheme
one would expect a mass difference of a few MeV between the two
neutral states. We note that in \cite{Maiani:2007vr} a strategy to
find the two missing charged partners is delineated. We refer to
this paper for further details.

\renewcommand{\arraystretch}{1.3}
\begin{table}[ht!]
\caption{\small \label{T1}Four-quark states with hidden charm
interpreted as bound states comprising a diquark ($cq$) and an
antidiquark ($\bar c\bar q$). $\{QQ\}$ (resp. $[QQ]$) means a spin 1
(resp. spin 0) diquark $QQ$. The masses of the present paper are
obtained by the Salpeter equation (see text).  The model in
\cite{Ebert:2005nc,Ebert:2007rn} uses a quasipotential of the
Schr\"odinger type \cite{Ebert:2002pp}. Ref. \cite{Maiani:2004vq}
uses a constituent quark model ($^\dagger$ means that the
experimental value is used as an input in this case). Units are GeV.
\label{diq}}\begin{center}
\begin{tabular}{|c|c|c|c|c|c|c|}\hline
 $J^{PC}$&Flavor content& Mass (this paper)& Mass
 \cite{Ebert:2005nc,Ebert:2007rn}&Mass \cite{Maiani:2004vq}&
  Exp. State &Exp Mass \\
 \hline $0^{++}$&$[cq][\bar c\bar q]$&3.857&3.812& 3.723&& \\
 \hline $1^{++}$&$([cq]\{\bar c\bar q\}+[\bar c\bar q]\{cq\})/\sqrt2$
 &3.899&3.871& $3.872^\dagger$& $X(3872)$&$3.8712\pm0.0004$
 \cite{Choi:2003ue,Abazov:2004kp,Acosta:2003zx,Aubert:2004ns,Abe:2005ix,Abe:2005iya}\\
 \hline $1^{+-}$&$([cq]\{\bar c\bar q\}-[\bar c\bar q]\{cq\})/\sqrt2$
 &3.899&3.871& $3.754$& &\\
 \hline $0^{++}$&$\{cq\}\{\bar c\bar q\}$&3.729&3.852& 3.832& &\\
 \hline $1^{+-}$&$\{cq\}\{\bar c\bar q\}$&3.833&3.890& 3.882& &\\
 \hline $2^{++}$&$\{cq\}\{\bar c\bar q\}$&3.988&3.968& 3.952& $
 Y(3940)$&$
 3.943\pm0.011\pm0.013$ \cite{Abe:2004zs}\\
\hline
\end{tabular}
\end{center}
\end{table}

\renewcommand{\arraystretch}{1.3}
\begin{table}[ht!]
\caption{\small \label{T2}Four-quark states with hidden bottom
interpreted as bound states comprising a diquark ($bq$) and an
antidiquark ($\bar b\bar q$). $\{QQ\}$ (resp. $[QQ]$) means a spin 1
(resp. $S=0$) diquark $QQ$. The masses of the present paper are
obtained by the Salpeter equation (see text).  The model in
\cite{Ebert:2005nc,Ebert:2007rn} uses a quasipotential of the
Schr\"odinger type \cite{Ebert:2002pp}.  Units are GeV.
\label{diqb}}\begin{center}
\begin{tabular}{|c|c|c|c|}\hline
 $J^{PC}$&Flavor content& Mass (this paper)& Mass
 \cite{Ebert:2005nc,Ebert:2007rn}\\
 \hline $0^{++}$&$[bq][\bar b\bar q]$&10.260&10.471 \\
 \hline $1^{+\pm}$&$([bq]\{\bar b\bar q\}\pm[\bar b\bar q]\{bq\})/\sqrt2$
 &10.284&10.492\\
 \hline $0^{++}$&$\{bq\}\{\bar b\bar q\}$&10.264&10.473\\
 \hline $1^{+-}$&$\{bq\}\{\bar b\bar q\}$&10.275&10.484\\
 \hline $2^{++}$&$\{bq\}\{\bar b\bar q\}$&10.296&10.534\\
\hline
\end{tabular}
\end{center}
\end{table}

By our model we can compute also radial excitations of tetraquarks.
For example the first radial excitations of the two $X$ states with
$1^{+-}$ have mass $m=4.421$ GeV and  $m=4.418$ GeV respectively.
 In Ref. \cite{Maiani:2007wz} the state $Z(4433)$, recently observed by
the Belle Collaboration \cite{:2007wga} through the decay
$Z(4433)\to \psi(2S)\pi^\pm$, is interpreted as the first radial
excitation of one of these states. Although the difference between
theory and experiment in this case is larger than, say, for
$X(3872)$ or $Y(3940)$, this interpretation is compatible with our
results because of the theoretical errors of the present model.
Another exotic state is $Y(4260)$, found  by the BaBar Collaboration
\cite{Aubert:2005rm} and confirmed by CLEO \cite{He:2006kg} and
BELLE \cite{:2007sj}. It is interpreted in \cite{Maiani:2005pe} as
an orbital excitation of a tetraquark state, an interpretation we
are  neither able to confirm nor to disprove, as we have limited the
analysis to the $\ell=0$ states.

Let us now comment on the light tetraquarks. We do not include them
in the Tables because for them the assumptions of our model and
therefore its results are less reliable; moreover, as discussed in
connections with  Eqns. (\ref{n1}), (\ref{n2}), the distortion of
the diquark wavefunction due to the relativistic motion is larger
for light diquarks and  its neglect generates a greater error.

\begin{table}[ht!]
\caption{\small \label{tetds} Comparison between the results of the
present model and those of Ref. \cite{Maiani:2004vq} for tetraquarks
with open charm and strangeness. Units are GeV.}
\begin{center}
\begin{tabular}{|c|c|c|c|}\hline
 $J^{P}$&Flavor content& Th. mass (this paper) & Th. mass (model
 \cite{Maiani:2004vq})\\
 \hline $0^{+}$&$[cq][\bar q\bar s]$&2.840& 2.371\\
\hline $0^{+}$&$\{cq\}\{\bar q\bar s\}$&2.503& 2.424\\
 \hline $1^{+}$&$\{cq\}[\bar q\bar s]$ & 2.880&  2.410\\
  \hline $1^{+}$&$\{cq\}\{\bar q\bar s\}$&2.748& 2.462\\
 \hline $1^{+}$&$[cq]\{\bar q\bar s\}$ & 2.841& 2.571 \\
 \hline $2^{+}$&$\{cq\}\{\bar q\bar s\}$&2.983& 2.648\\
\hline
\end{tabular}
\end{center}
\end{table}

Let us finally comment on the  possible existence of tetraquarks
comprising a heavy diquark and a light diquark. We present in Table
\ref{tetds} our predictions for tetraquarks with open charm and
 strangeness and compare them with the prediction of the
constituent quark model of Ref. \cite{Maiani:2004vq}. In
\cite{Maiani:2004vq} the state $0^+$ is associated with the particle
$D_s(2317)$ \cite{Aubert:2003fg}, $1^+$ with $D_s(2457)$
\cite{Aubert:2003fg} and $2^+$ with $X(2632)$
\cite{Evdokimov:2004iy}. Our results are significantly different
from those of \cite{Maiani:2004vq}. Again, this might be due to the
limitations of one or both the constituent quark models. In any
event, on the basis of the results in Section \ref{III}  we do not
expect theoretical errors larger than a few hundred MeV for the
results of the present model in Table \ref{tetds}, so that we do not
support the interpretation of the states $D_s(2317)$, $D_s(2457)$
 and  $X(2632)$ as tetraquark charmed states with open strangeness.
 In \cite{Vijande:2006hj} these states are interpreted as a mixture of $P$-wave quark-antiquark
  states and four-quark components.

\section{\label{VII}Conclusions}
We have developed an application and, at the same time, a test for
the QCD potential found by means of the AdS/QCD correspondence. We
have put it in a semirelativistic wave equation and fitted meson
spectra. Our result is that this model, with the AdS/QCD potential
plus a contribution from spin interaction, can reproduce
 the experimental data except for the lighter states
($\pi,\,K$).  This agreement has motivated us to make some
predictions on the masses of tetraquarks, considering them as bound
states of a diquark and an antiquark. Our conclusions are that some
tetraquark states with appropriate flavor content can be identified
with the particles $X(3872)$ and $Y(3940)$. On the other hand the
present model does not favor the interpretation of some charmed
positive parity particles with strangeness as tetraquark
states.

 \vspace{0.2truecm}
 \textbf{Acknowledgements}

 We thank P.Colangelo,  F. De Fazio and S. Nicotri for useful
 discussions and R. Jaffe, L. Maiani, A. Polosa and F. Sch\"oberl for valuable correspondence.


\vskip.1cm


\begin{thebibliography}{99}

\bibitem{Ricciardi:2007rs}
  S.~Ricciardi  [BaBar Collaboration],
  AIP Conf.\ Proc.\  {\bf 892}, 456 (2007).
\bibitem{Zupanc:2007rw}
  A.~Zupanc  [Belle Collaboration],
  AIP Conf.\ Proc.\  {\bf 892}, 472 (2007).
\bibitem{Swanson:2006st}
  E.~S.~Swanson,
  Phys.\ Rept.\  {\bf 429}, 243 (2006)
  [arXiv:hep-ph/0601110].
\bibitem{Jaffe:2004ph}
  R.~L.~Jaffe,
  Phys.\ Rept.\  {\bf 409}, 1 (2005)
  [Nucl.\ Phys.\ Proc.\ Suppl.\  {\bf 142}, 343 (2005)]
  [arXiv:hep-ph/0409065].
\bibitem{Choi:2003ue}
  S.~K.~Choi {\it et al.}  [Belle Collaboration],
  Phys.\ Rev.\ Lett.\  {\bf 91}, 262001 (2003)
  [arXiv:hep-ex/0309032].
\bibitem{Acosta:2003zx}
  D.~Acosta {\it et al.}  [CDF II Collaboration],
  Phys.\ Rev.\ Lett.\  {\bf 93}, 072001 (2004)
  [arXiv:hep-ex/0312021].
\bibitem{Abazov:2004kp}
  V.~M.~Abazov {\it et al.}  [D0 Collaboration],
  Phys.\ Rev.\ Lett.\  {\bf 93}, 162002 (2004)
  [arXiv:hep-ex/0405004].
\bibitem{Aubert:2004ns}
  B.~Aubert {\it et al.}  [BABAR Collaboration],
  Phys.\ Rev.\  D {\bf 71}, 071103 (2005)
  [arXiv:hep-ex/0406022].
\bibitem{Abe:2005ix}
  K.~Abe {\it et al.},
  arXiv:hep-ex/0505037.
\bibitem{Abe:2005iya}
  K.~Abe {\it et al.},
  arXiv:hep-ex/0505038.
\bibitem{Abe:2004zs}
  K.~Abe {\it et al.}  [Belle Collaboration],
  Phys.\ Rev.\ Lett.\  {\bf 94}, 182002 (2005)
  [arXiv:hep-ex/0408126].
\bibitem{Maiani:2004uc}
  L.~Maiani, F.~Piccinini, A.~D.~Polosa and V.~Riquer,
  Phys.\ Rev.\ Lett.\  {\bf 93}, 212002 (2004)
  [arXiv:hep-ph/0407017].
  %
\bibitem{Maiani:2004vq}
L.~Maiani, F.~Piccinini, A.~D.~Polosa and V.~Riquer,
  Phys.\ Rev.\  D {\bf 71}, 014028 (2005)
  [arXiv:hep-ph/0412098].
  %
\bibitem{Maiani:2005pe}
  L.~Maiani, V.~Riquer, F.~Piccinini and A.~D.~Polosa,
  Phys.\ Rev.\  D {\bf 72}, 031502(R) (2005)
  [arXiv:hep-ph/0507062].
\bibitem{Maiani:2007vr}
  L.~Maiani, A.~D.~Polosa and V.~Riquer,
  arXiv:0707.3354 [hep-ph].
\bibitem{De Rujula:1975ge}
  A.~De Rujula, H.~Georgi and S.~L.~Glashow,
  Phys.\ Rev.\  D {\bf 12}, 147 (1975).

\bibitem{Colangelo:1990rv}
  P.~Colangelo, G.~Nardulli and M.~Pietroni,
  Phys.\ Rev.\  D {\bf 43}, 3002 (1991).

\bibitem{Eichten:1978tg}
  E.~Eichten, K.~Gottfried, T.~Kinoshita, K.~D.~Lane and T.~M.~Yan,
  Phys.\ Rev.\  D {\bf 17}, 3090 (1978)
  [Erratum-ibid.\  D {\bf 21}, 313 (1980)].
\bibitem{Richardson:1978bt}
  J.~L.~Richardson,
  Phys.\ Lett.\  B {\bf 82}, 272 (1979).
\bibitem{Buchmuller:1980su}
  W.~Buchmuller and S.~H.~H.~Tye,
  Phys.\ Rev.\  D {\bf 24}, 132 (1981).

\bibitem{Andreev:2006ct}
  O.~Andreev and V.~I.~Zakharov,
  Phys.\ Rev.\  D {\bf 74}, 025023 (2006)
  [arXiv:hep-ph/0604204].


\bibitem{Aubert:2003fg}
  B.~Aubert {\it et al.}  [BABAR Collaboration],
  Phys.\ Rev.\ Lett.\  {\bf 90}, 242001 (2003)
  [arXiv:hep-ex/0304021].
\bibitem{Evdokimov:2004iy}
  A.~V.~Evdokimov {\it et al.}  [SELEX Collaboration],
  Phys.\ Rev.\ Lett.\  {\bf 93}, 242001 (2004)
  [arXiv:hep-ex/0406045].
\bibitem{ads/cft}
J. Maldacena, Adv.Theor.Math.Phys. 2, 231 (1998).
\bibitem{malda}
J. Maldacena, Phys.Rev.Lett. 80, 4859 (1998); S.-J. Rey and J.-T.
Yee, Eur.Phys.J. C22, 379 (2001).
\bibitem{ADSCFT}
The following is an incomplete list: E. Witten, Adv. Theor. Math.
Phys. 2 505 (1998) ; J. A. Minahan and N. P. Warner, JHEP 9806 005 (1998)
; H. Dorn and H.-J. Otto, JHEP 9809, 021 (1998); J. Greensite and
P. Olesen, JHEP 001  9904 (1999), ; N. Drukker, D.J. Gross, and H.
Ooguri, Phys.Rev. D {\bf 60} 125006 (1999); A. M. Polyakov and V. S.
Rychkov, Nucl.Phys. B581 116 (2000).

\bibitem{Erlich:2005qh}
  J.~Erlich, E.~Katz, D.~T.~Son and M.~A.~Stephanov,
  Phys.\ Rev.\ Lett.\  {\bf 95}, 261602 (2005)
  [arXiv:hep-ph/0501128].
\bibitem{Karch:2006pv}
  A.~Karch, E.~Katz, D.~T.~Son and M.~A.~Stephanov,
  Phys.\ Rev.\  D {\bf 74}, 015005 (2006)
  [arXiv:hep-ph/0602229].
\bibitem{Andreev:2006vy}
  O.~Andreev,
  Phys.\ Rev.\  D {\bf 73}, 107901 (2006)
  [arXiv:hep-th/0603170].
\bibitem{Cea:1986bj}
  P.~Cea and G.~Nardulli,
  Phys.\ Rev.\  D {\bf 34}  1863 (1986).
  \bibitem{Barnes:2005pb}
  T.~Barnes, S.~Godfrey and E.~S.~Swanson,
  Phys.\ Rev.\  D {\bf 72}, 054026 (2005)
  [arXiv:hep-ph/0505002].
  \bibitem{Karamcheti}K. Karamcheti,
  {\it Principles of Ideal Fluid Aerodynamics} (New York, 1966).
  \bibitem{PDG} W.-M. Yao et al. [Particle Data Group], J. Phys. G 33, 1 (2006)
\bibitem{Ebert:2005nc}
  D.~Ebert, R.~N.~Faustov and V.~O.~Galkin,
  Phys.\ Lett.\  B {\bf 634}, 214 (2006)
  [arXiv:hep-ph/0512230].
\bibitem{Ebert:2007rn}
  D.~Ebert, R.~N.~Faustov, V.~O.~Galkin and W.~Lucha,
  arXiv:0706.3853 [hep-ph].

\bibitem{Ebert:2002pp}
  D.~Ebert, R.~N.~Faustov and V.~O.~Galkin,
  Phys.\ Rev.\  D {\bf 67}, 014027 (2003)
  [arXiv:hep-ph/0210381].

\bibitem{Jaffeold}
  R.~L.~Jaffe,
  Phys.\ Rev.\  D {\bf 15}, 281 (1977);
  R.~L.~Jaffe and F.~E.~Low,
  Phys.\ Rev.\  D {\bf 19}, 2105 (1979);
  M.~G.~Alford and R.~L.~Jaffe,
  Nucl.\ Phys.\  B {\bf 578}, 367 (2000)
  [arXiv:hep-lat/0001023].
%
\bibitem{Tornqvist:2004qy}
  N.~A.~Tornqvist,
  Phys.\ Lett.\  B {\bf 590}, 209 (2004)
  [arXiv:hep-ph/0402237].

\bibitem{Swanson:2003tb}
  E.~S.~Swanson,
  Phys.\ Lett.\  B {\bf 588}, 189 (2004)
  [arXiv:hep-ph/0311229].

\bibitem{Close:2003sg}
  F.~E.~Close and P.~R.~Page,
  Phys.\ Lett.\  B {\bf 578}, 119 (2004)
  [arXiv:hep-ph/0309253].
\bibitem{Pakvasa:2003ea}
  S.~Pakvasa and M.~Suzuki,
  Phys.\ Lett.\  B {\bf 579}, 67 (2004)
  [arXiv:hep-ph/0309294].
\bibitem{Wong:2003xk}
  C.~Y.~Wong,
  Phys.\ Rev.\  C {\bf 69}, 055202 (2004)
  [arXiv:hep-ph/0311088].
\bibitem{Braaten:2004rw}
  E.~Braaten and M.~Kusunoki,
  Phys.\ Rev.\  D {\bf 69}, 114012 (2004)
  [arXiv:hep-ph/0402177].
\bibitem{Barnes:2003vb}
  T.~Barnes and S.~Godfrey,
  Phys.\ Rev.\  D {\bf 69}, 054008 (2004)
  [arXiv:hep-ph/0311162].
  \bibitem{Li:2004sta}
  B.~A.~Li,
  Phys.\ Lett.\  B {\bf 605}, 306 (2005)
  [arXiv:hep-ph/0410264].
\bibitem{Hogaasen:2005jv}
  H.~Hogaasen, J.~M.~Richard and P.~Sorba,
  Phys.\ Rev.\  D {\bf 73}, 054013 (2006)
  [arXiv:hep-ph/0511039].
 \bibitem{Bernard:1997ib}
  C.~W.~Bernard, {\it et al.}  [MILC Collaboration],
  \emph{Phys.\ Rev.\ D } {\bf 56}, 7039 (1997)
  [arXiv:hep-lat/9707008].
\bibitem{Mei:2002ip}
  Z.~H.~Mei, and X.~Q.~Luo,
     \emph{Int.\ J.\ Mod.\ Phys.\ A} {\bf 18}, 5713 (2003)
  [arXiv:hep-lat/0206012].
\bibitem{Abe:2005hd}
  K.~Abe {\it et al.},
  arXiv:hep-ex/0507019.
  \bibitem{Maiani:2007wz}
  L.~Maiani, A.~D.~Polosa and V.~Riquer,
  arXiv:0708.3997 [hep-ph].


\bibitem{:2007wga}
  K.~Abe {\it et al.}  [Belle Collaboration],
  arXiv:0708.1790 [hep-ex].





\bibitem{Aubert:2005rm}
  B.~Aubert {\it et al.}  [BABAR Collaboration],
  Phys.\ Rev.\ Lett.\  {\bf 95}, 142001 (2005)
  [arXiv:hep-ex/0506081].
\bibitem{He:2006kg}
  Q.~He {\it et al.}  [CLEO Collaboration],
  Phys.\ Rev.\  D {\bf 74}, 091104 (2006)
  [arXiv:hep-ex/0611021].
\bibitem{:2007sj}
  C.~Z.~Yuan {\it et al.}  [Belle Collaboration],
  arXiv:0707.2541 [hep-ex].

\bibitem{Vijande:2006hj}
  J.~Vijande, F.~Fernandez and A.~Valcarce,
  Phys.\ Rev.\  D {\bf 73} 034002 (2006)
  [Erratum-ibid.\  D {\bf 74} 059903 (2006)]
  [arXiv:hep-ph/0601143].







  \end{thebibliography}
\end{document}